%% file: main.tex
\documentclass[aps,showpacs,pra,superscriptaddress,preprintnumbers,amsmath,amssymb,twocolumn,nolongbibliography]{revtex4-2}
\usepackage[utf8]{inputenc}
\usepackage[T1]{fontenc}
\usepackage{graphicx}
\usepackage{dcolumn}
\usepackage{appendix}
\usepackage{amsmath}
\usepackage{bm}
\usepackage[dvipsnames]{xcolor}
\usepackage{multirow}
\usepackage{notes2bib}
\usepackage{titletoc}
\usepackage{lipsum}
\usepackage{accents}

\usepackage{subfigure}
\usepackage{bbm}
\usepackage{enumerate}
\usepackage[
bookmarks=true,
colorlinks,
linkcolor=black,
urlcolor=black,
citecolor=black,
plainpages=false,
pdfpagelabels,
final,
breaklinks=true
]{hyperref}
\usepackage{titlesec}
\usepackage{array}
\usepackage{physics}

\newcommand{\relaxket}[1]{\lvert{#1}\rangle}

\begin{document}
\allowdisplaybreaks

\title{Generation of circular polarized high-order harmonics from single color quantum light}

\author{L.~Petrovic}
\email{lidija.petrovic@icfo.eu}
\affiliation{ICFO -- Institut de Ciencies Fotoniques, The Barcelona Institute of Science and Technology, 08860 Castelldefels (Barcelona)}

\author{P.~Stammer}
\affiliation{ICFO -- Institut de Ciencies Fotoniques, The Barcelona Institute of Science and Technology, 08860 Castelldefels (Barcelona)}
\affiliation{Atominstitut, Technische Universität Wien, 1020 Vienna, Austria}

\author{M. Lewenstein}
\affiliation{ICFO -- Institut de Ciencies Fotoniques, The Barcelona Institute of Science and Technology, 08860 Castelldefels (Barcelona)}

\author{J.~Rivera-Dean}
\email{physics.jriveradean@proton.me}
\affiliation{ICFO -- Institut de Ciencies Fotoniques, The Barcelona Institute of Science and Technology, 08860 Castelldefels (Barcelona)}
\affiliation{Department of Physics and Astronomy, University College London, Gower Street, London WC1E 6BT, UK}

\begin{abstract}
	The atomic response to an ultra-intense driving field produces a characteristic high-harmonic spectrum featuring a rapid drop in intensity for the lower harmonics, followed by a plateau and a sharp cutoff. This response vanishes for circularly polarized classical drivers---a limitation that can be overcome by introducing quantum features into the driving field.~In this work, we show that squeezed highly elliptically polarized drivers not only enable the high-harmonic generation (HHG) process in classically forbidden regimes of large ellipticity, but also yield highly elliptical harmonic radiation with pronounced super-Poissonian photon statistics.~Moreover, we show that the HHG spectral features encode information about the quantum nature of the driving field, revealing the presence of its squeezed field fluctuations.~By analyzing the HHG spectral intensity dependence as a function of the driver's ellipticity and squeezing orientation, we identify a means to probe the driving field's quantum properties that intrinsically lie in the high-photon number regime.
\end{abstract}

\maketitle

\emph{Introduction.---}In recent years, there has been a growing interest in the intersection of quantum optics and strong-field physics~\cite{lamprou_generation_2024,stammer2025colloquium, rivera_new_2025}, giving rise to what is often called \emph{Extreme Quantum Optics} or \emph{Strong-Field Quantum Optics}. This nascent field seeks to combine strengths from strong-field physics, such as the generation of ultrafast and broadband radiation via high-harmonic generation (HHG)~\cite{antoine_attosecond_1996,drescher_x-ray_2001,paul_observation_2001}, with quantum optics techniques that exploit the properties of non-classical light~\cite{gisin_quantum_2007,kok_linear_2007,giovannetti_advances_2011,aspuru-guzik_photonic_2012,pirandola_advances_2018}. These combined approaches could open new avenues for ultrafast strong-field quantum optical applications~\cite{stammer_metrological_2024,cruz-rodriguez_quantum_2024,sennary_ultrafast_2025}.~In this direction, recent developments have demonstrated the potential of this connection for generating high-photon number non-classical states of light~\cite{lewenstein_generation_2021}, involving, and possibly entangling, a broad range of spectral modes~\cite{stammer_entanglement_2024,rivera-dean_squeezed_2024,rivera-dean_about_2024,lange_electron-correlation-induced_2024,theidel_evidence_2024,theidel_observation_2024,lange_hierarchy_2024,lange_excitonic_2025,yi_generation_2024,stammer_theory_2025,rivera-dean_attosecond_2025}.

Equally important, however, is understanding how each of these areas benefit individually from this new connection.~Recently, the generation of bright squeezed light~\cite{spasibko_multiphoton_2017,manceau_indefinite-mean_2019} with parameters sufficient to drive strong-field interactions~\cite{lemieux_photon_2024,rasputnyi_high_2024,heimerl_multiphoton_2024,tzur_measuring_2025,heimerl_driving_2025} has been proposed to markedly modify the HHG properties~\cite{gorlach_high-harmonic_2023,rivera-dean_structured_2025,stammer_weak_2025}, and nontrivially alter its sub-cycle electron dynamics~\cite{even_tzur_photon-statistics_2023,rivera-dean_structured_2025, stammer_weak_2025}.~These advances have enabled new routes to both manipulate~\cite{tzur_generation_2023,lemieux_photon_2024,stammer_weak_2025,rivera-dean_attosecond_2025} and characterize~\cite{tzur_measuring_2025,rivera-dean_attosecond_2025} the emitted radiation in regimes inaccessible with linear optics~\cite{lamprou_generation_2024}.~A particularly illustrative example, central to this work, is the generation of harmonic radiation under otherwise classically forbidden conditions, such as circularly polarized driving fields~\cite{budil_influence_1993,dietrich_high-harmonic_1994,corkum_subfemtosecond_1994,liang_experimental_1995,burnett_ellipticity_1995,antoine_theory_1996}.~Combining squeezed polarization components in circularly polarized configurations can profoundly modify the ultrafast electron dynamics, thereby re-enabling the HHG process~\cite{rivera-dean_structured_2025}.

\begin{figure}
    \centering
    \includegraphics[width=1\columnwidth]{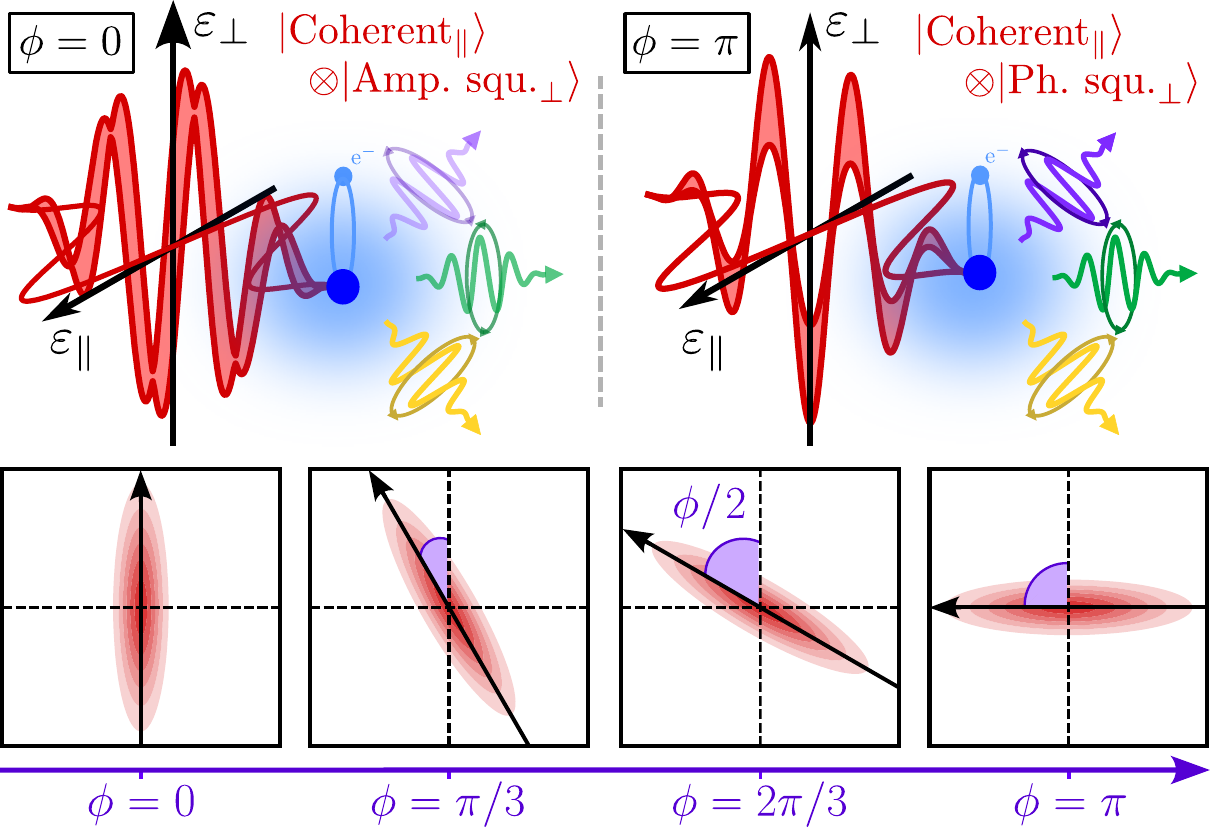}
    \caption{Schematic illustration of the use of bright squeezed elliptically polarized light to characterize the driving field properties and generate highly elliptically polarized harmonic radiation.~We consider configurations where one linear polarization component (here the $\perp$-component) is a displaced squeezed vacuum state, while the other (the $\parallel$-component) remains a coherent state. By varying the the driver's ellipticity and consequently the squeezing orientation, specified by $\phi$, one can control the intensity, ellipticity, and photon statistics of the emitted radiation.
    }
    \label{Fig:Scheme}
\end{figure}

In this Letter, we leverage HHG driven by elliptically polarized squeezed light as a versatile tool to both characterize the driving field properties and generate highly elliptically polarized harmonic radiation using single-color drivers [Fig.~\ref{Fig:Scheme}].~First, we demonstrate that the sensitivity of the harmonic response to the type of squeezing~\cite{rivera-dean_structured_2025} can serve to certify the field fluctuations present in the strong-field squeezed driver.~Second, we analyze the characteristics of the emitted harmonic radiation.~Specifically, we show that this approach enables the generation of highly elliptically polarized harmonics---a feature that, classically, requires engineered bichromatic fields~\cite{fleischer_spin_2014,pisanty_spin_2014}.~Finally, we examine the photon statistics properties of the generated harmonics.~We report the presence of pronounced super-Poissonian behavior, which we attribute to the interplay between the squeezing-induced field fluctuations and the strong nonlinearity of the HHG process.

More broadly, this work highlights the reciprocal benefits between quantum optics and strong-field physics. Quantum optics provides new tools for strong-field physics by controlling and characterizing strong-field phenomena, here enabling the generation of highly elliptical polarized harmonics using single-color fields.~Conversely, the strong-field response to high-intensity quantum light opens new avenues for probing non-classical fields in the domain where strong-field physics naturally operates: the extreme wavelength, and high-photon number regime.


	
	

\emph{Characterizing the driving field properties.---}\input{Driver}

\emph{Elliptically polarized harmonics.---}\input{Ellipticity}

\emph{Photon statistics of the harmonics.---}\input{PhotonStatistics} 

\emph{Conclusions.---}This work has explored the versatility of HHG driven by elliptically polarized fields with strong squeezing features: it serves as a sensitive probe of field fluctuations in the driving light and as a source of highly elliptically polarized harmonic radiation.~In particular, we have shown that the generated harmonics are strongly influenced by asymmetric field fluctuations in the driver, with both the squeezing angle and the driving-field ellipticity as tunable parameters to control modifications on the generated harmonics~[Fig.~\ref{Fig:Driver:Charac}].~Moreover, we have demonstrated the generation of highly elliptically polarized harmonic light under nearly circularly polarized single-color driving fields~[Fig.~\ref{Fig:HHG:Ellipticity}], and shown that such emission exhibits markedly super-Poissonian photon statistics, with extremely large values of the second-order autocorrelation function~[Fig.~\ref{Fig:g2:Amp:squ}].~This behavior arises from the interplay between the large quantum fluctuations characteristic of strongly squeezed light and the highly nonlinear nature of the HHG process.

On a broader scale, our work demonstrates how HHG driven by elliptically polarized squeezed light can benefit both the quantum optics and strong-fields communities: for the former, as a diagnostic tool of non-classical features in high-photon number regimes; and for the latter, as a novel means of generating highly elliptically polarized harmonics using single-color drivers. Regarding the latter, we find that this configuration does not yield near-coherent harmonic radiation. However, our results suggest that strong squeezing could partially overcome this limitation by inducing significant atomic depletion, which on the other hand has been recently shown to significantly influence the propagation of the emitted radiation~\cite{rivera-dean_propagation_2025}. Similar effects have been found to generate squeezing and entanglement among different harmonic orders even for coherent state drivers~\cite{stammer_entanglement_2024,lange_hierarchy_2024}.~Understanding how these mechanisms modify the quantum optical properties of the harmonics will be essential to fully assess the potential of elliptically polarized, squeezed-light-driven HHG both as a probe of non-classical light and as a source of highly elliptically polarized radiation.


\emph{Acknowledgments.---}J.~R.-D.~is grateful to Misha Ivanov for valuable discussions that inspired part of the results reported in this work, and to Marcin Płodzień for all the support and interest shown throughout its development.

ICFO-QOT group acknowledges support from: MCIN/AEI (PGC2018-0910.13039/501100011033,  CEX2019-000910-S/10.13039/501100011033, Plan National STAMEENA PID2022-139099NB, project funded MCIN and  by the ``European Union NextGenerationEU/PRTR'' (PRTR-C17.I1), FPI); Ministry for Digital Transformation and of Civil Service of the Spanish Government through the QUANTUM ENIA project call - Quantum Spain project, and by the European Union through the Recovery, Transformation and Resilience Plan - NextGenerationEU within the framework of the Digital Spain 2026 Agenda; CEX2024-001490-S [MICIU/AEI/10.13039/501100011033]; Fundació Cellex; Fundació Mir-Puig; Generalitat de Catalunya (European Social Fund FEDER and CERCA program; Barcelona Supercomputing Center MareNostrum (FI-2023-3-0024); Funded by the European Union (HORIZON-CL4-2022-QUANTUM-02-SGA, PASQuanS2.1, 101113690, EU Horizon 2020 FET-OPEN OPTOlogic, Grant No 899794, QU-ATTO, 101168628),  EU Horizon Europe Program (No 101080086 NeQSTGrant Agreement 101080086 — NeQST).

P. Stammer acknowledges support from: The European Union’s Horizon 2020 research and innovation programme under the Marie Skłodowska-Curie grant agreement No 847517.

\bibliography{References.bib}{}

\include{Supplementary_Material}

\end{document}

%% file: Driver.tex
In this work, we focus on HHG driven by non-classical states of the form
\begin{equation}\label{Eq:init:state}
	\ket{\psi}
		= \ket{\bar{\alpha}}_{\parallel}
				\otimes \hat{D}_{\perp}(i A \bar{\alpha})
				\otimes \hat{S}_{\perp}(\xi)\ket{0}_{\perp},
\end{equation}
where one polarization component $(\parallel)$ is prepared in a coherent state of amplitude $\bar{\alpha}$, while the orthogonal mode ($\perp$) is in a displaced squeezed vacuum (DSV) state. The DSV is defined through a displacement operator $\hat{D}_{\mu}(\alpha) = \exp[\alpha \hat{a}_{\mu}^\dagger - \alpha^* \hat{a}_{\mu}]$, and the squeezing operator $\hat{S}_\mu(\xi) = \exp[\xi^* \hat{a}_\mu^2 - \xi \hat{a}_\mu^{\dagger 2}]$ acting on a vacuum state, with $\hat{a}_\mu$ ($\hat{a}^\dagger_\mu$) the annihilation (creation) operator for the $\mu$-polarization mode. In Eq.~\eqref{Eq:init:state}, the parameter $A\in[0,1]$ controls the ellipticity of the coherent component, while the squeezing parameter $\xi = r e^{-i\phi}$ introduces quantum features in the driver.~For $\xi = 0$, the state reduces to a classical linearly polarized field when $A=0$, and to a circularly polarized field when $A=1$. Introducing $r>0$ adds squeezing along the $\perp$-polarization, with $\phi$ rotating the quadrature that gets squeezed: $\phi = 0$ results in amplitude squeezing, whereas $\phi = \pi$ yields phase squeezing [Fig.~\ref{Fig:Scheme} upper panels]. Thus, for a fixed squeezing amplitude $r$, the role of $\phi$ is to effectively rotate the squeezed state in phase space around its origin [Fig.~\ref{Fig:Scheme} lower panels].~Importantly, adding squeezing to elliptically polarized drivers leads to a modified polarization with an effective ellipticity, especially in regimes of strong squeezing~\cite{rivera-dean_structured_2025}.

An essential requirement in Eq.~\eqref{Eq:init:state} is that the squeezing is strong enough to significantly perturb the HHG process.~The intensity of a DSV state can be decomposed as $I_{\text{DSV}} = I_{\text{coh}} + I_{\text{sq}}$, where $I_{\text{coh}} \equiv \epsilon^2 \abs{\bar{\alpha}}$ and $I_{\text{sq}} \equiv \epsilon^2 \sinh[2](r)$ correspond to the contributions from the coherent and squeezed components, respectively.~Here, $\epsilon = \sqrt{\hbar \omega_L/(2\epsilon_0 V)}$ denotes the light-matter coupling, with $V$ the quantization volume.~Thereby, the condition stated above is typically satisfied when $I_{\text{coh}}/I_{\text{sq}} \sim 10^{-2}$, values achievable with state-of-the-art high-gain spontaneous parametric down-conversion~\cite{spasibko_multiphoton_2017,manceau_indefinite-mean_2019}. Squeezed states of this type have already been employed to induce tunneling ionization in metal needle tips~\cite{heimerl_driving_2025}, to drive HHG in semiconductor materials~\cite{rasputnyi_high_2024}, and as $\omega$-perturbations assisting HHG initiated by an intense classical $(\omega/2)$-field in solids~\cite{lemieux_photon_2024}, and a $2\omega$-field in gases~\cite{tzur_measuring_2025}. In what follows, we consider conditions where the coherent component has an electric field strength $\bar{\varepsilon} = 0.053$ a.u., where $\bar{\varepsilon} = 2 \epsilon \bar{\alpha}$, and $I_{\text{sq}} = 10^{-5}$ a.u. for the squeezing contribution, allowing the composed field to both classically drive and non-classically modify the HHG process in atomic media.~Hereupon, we focus on hydrogen, with ionization potential $I_p = 0.5$ a.u., as the HHG target system.

Under the conditions described above, and for $A=1$ in Eq.~\eqref{Eq:init:state} corresponding to a circular polarized field, such configurations enable the generation of HHG radiation across several harmonic orders~\cite{rivera-dean_structured_2025}---a process that becomes increasingly suppressed in the classical regime without squeezing as the driver's ellipticity grows~\cite{budil_influence_1993,antoine_theory_1996}. Of particular relevance here is that the HHG spectra are strongly shaped by the type of squeezing applied, with amplitude and phase squeezing yielding different HHG cutoffs dependencies with the squeezing intensity $I_{\text{sq}}$~\cite{rivera-dean_structured_2025} (see Supplementary Material~\ref{Sec:SM:TheoryBackground}).~This suggests the potential of HHG driven by squeezed circularly polarized light as a probe of the driver's non-classical properties, with the squeezing orientation $\phi$ as a control parameter. 

To gain further insight on this, we evaluate the HHG spectrum produced by Eq.~\eqref{Eq:init:state}, which can be written in general as~\cite{gorlach_high-harmonic_2023,rivera-dean_structured_2025} (see Supplementary Material~\ref{Sec:SM:TheoryBackground})
\begin{align}\label{Eq. Final generalized spectrum}
	S_{\text{sq}}(\omega,\phi,A)\!
		 \propto \!\!
		 	\int\! \dd \Tilde{\varepsilon}^{(\perp)}_{\alpha} 
		 		\mathcal{Q}_{\perp}(\Tilde{\varepsilon}^{(\perp)}_{\alpha})
				\big[
					|d^{(\parallel)}_{\boldsymbol{\varepsilon}_{\alpha}}(\omega)|^2 
					+d^{(\perp)}_{\boldsymbol{\varepsilon}_{\alpha}}|^2
				\big].
\end{align}
Here, $\mathcal{Q}_{\perp}(\Tilde{\varepsilon}_{\alpha,\perp})$ is a marginal of the Husimi function~\cite{SchleichBookCh12}, whose specific form encodes the non-classical properties of the driver, and $d_{\boldsymbol{\varepsilon}}^{(\mu)}(\omega)$ is the Fourier Transform of the semiclassical time-dependent dipole moment along the $\mu$-polarization when driven by a field of amplitude $\boldsymbol{\varepsilon} = (\varepsilon_{\parallel},\varepsilon_{\perp})$.~Calculations are performed in the classical limit, $V\to \infty$ (equivalently $\epsilon\to 0$) and $\alpha \to \infty$ such that $\varepsilon_{\alpha}$ remains finite, a suitable approach for strong-field problems with freely propagating fields containing an extremely large photon number~\cite{gorlach_high-harmonic_2023}.~Here, $\varepsilon_{\alpha}$ and $\tilde{\varepsilon}_{\alpha}$ are related through $\Tilde{\varepsilon}_{\alpha,x} = (\varepsilon_x - \bar{\varepsilon}_x)\cos(\phi/2) - (\varepsilon_y - \bar{\varepsilon}_y)\sin(\phi/2)$ and $\Tilde{\varepsilon}_{\alpha,y} = (\varepsilon_y - \bar{\varepsilon}_y)\cos(\phi/2) + (\varepsilon_x - \bar{\varepsilon}_x)\sin(\phi/2)$, with $\varepsilon_{\alpha} = \varepsilon_{\alpha,x} + i \varepsilon_{\alpha,y}$ (see Supplementary Material~\ref{Sec:SM:TheoryBackground}).

\begin{figure}
	\centering
	\includegraphics[width=1\columnwidth]{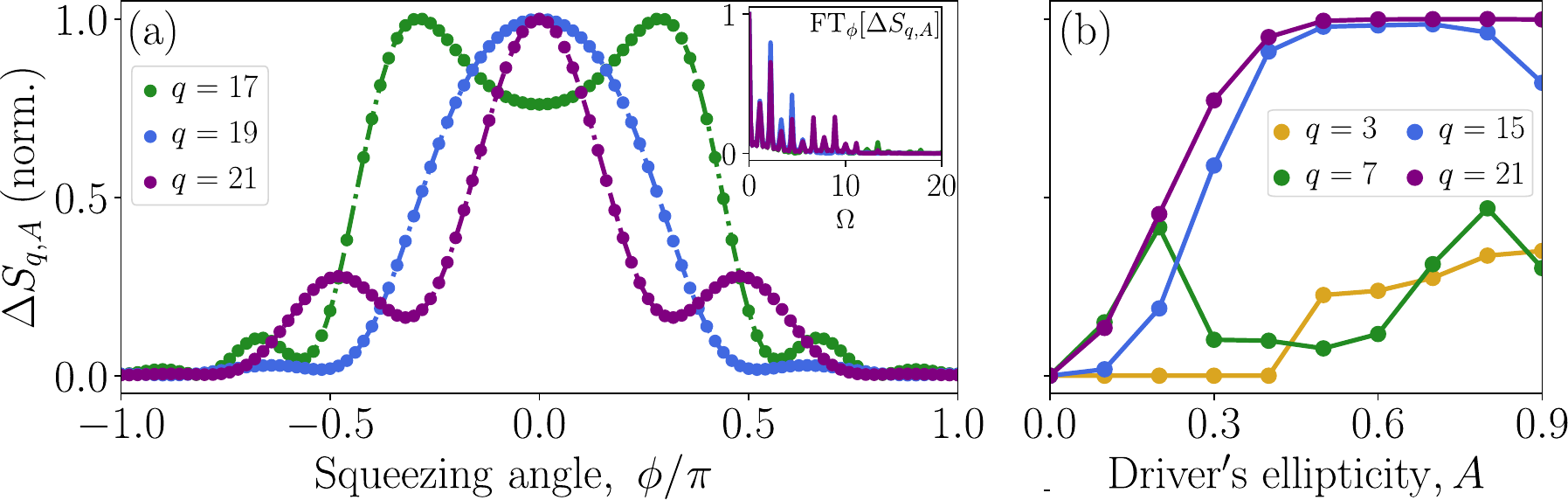}
	\caption{(a) Normalized intensity of harmonic orders $q = 17,19$ and $21$ against the squeezing angle $\phi$ for $A = 0.9$, with the inset plot showing the Fourier transform of the obtained signals. The FT is obtained by taking into account that the pattern in $\Delta S_{q,A}(\phi)$ repeats itself every $2\pi$ due to the recurring squeezing directions.~(b)~Normalized intensity difference between amplitude and phase squeezing directions for different harmonic orders against the driver's ellipticity $A$.~Calculations have been performed over five optical cycles of a monochromatic field with set $\overline{\varepsilon}_\parallel = 0.053$ a.u., $\omega_L = 0.057$ a.u., $I_{\text{sq},\perp} = 10^{-5}$ a.u., using $I_p = 0.5$ a.u.~(hydrogen) for the ionization potential.}
	\label{Fig:Driver:Charac}
\end{figure}

More precisely, we evaluate the normalized intensity difference $\Delta S_{q,A}(\phi) = \big| 1 - \frac{S_{\phi}(\omega_q)}{S_{\phi=\pi}(\omega_q)}\big|$ for harmonic orders $\omega_q = q \omega_L$ and $A = 0.9$, i.e., an almost circularly-polarized driver in the absence of squeezing.~In practice, this compares the $q$th harmonic intensity obtained for an elliptically polarized field with squeezing parameter $\xi = r e^{-i\phi}$ against that from phase squeezing ($\phi = \pi$), which overall yields the strongest signal (see Supplementary Material~\ref{Sec:SM:TheoryBackground}).

Figure~\ref{Fig:Driver:Charac}~(a) presents these results near the cutoff region, where intensity differences across varying $\phi$ are more pronounced.~As $\phi$ varies in the range $[-\pi,\pi]$, each harmonic order displays a characteristic pattern, with $\Delta S_{q,A}(\phi) = 0$ for phase squeezing by definition, and increasing sharply as $\phi$ approaches zero.~At this point, $\Delta S_{q,A}(\phi)$ quantifies the (normalized) intensity difference obtained between amplitude- and phase-squeezed driving fields.~Such enhancement at intermediate $\phi$ arises from how the squeezed field fluctuations map onto the HHG spectrum, and how this mapping can be tuned via $\phi$~\cite{rivera-dean_structured_2025}, that is, by rotating the squeezed state around its origin in phase space [Fig.~\ref{Fig:Scheme} lower panels].~This mechanism is particularly useful in the strong-field regime, where more conventional signatures---such as the photon-number probability distribution---are challenging or even impossible to measure accurately.~In particular, light sources that are homogeneous in phase space, such as coherent and thermal states, are invariant under a rotation $\phi$ in phase space [Fig.~\ref{Fig:Scheme} lower panels], making the features seen in Fig.~\ref{Fig:Driver:Charac} absent. Therefore, the presence of $\Delta S_{q,A}(\phi) \neq 0$, or equivalently non-zero frequencies in its Fourier transform [inset of Fig.~\ref{Fig:Driver:Charac}~(a)], make this quantity an HHG-based witness of asymmetric strong-field fluctuations. 

Interestingly, the squeezing direction is not the only handle for probing the anisotropy of the driving field fluctuations, but also its ellipticity~$A$.~Figure~\ref{Fig:Driver:Charac}~(b) shows $\Delta S_{q,A}(\phi=0)$, thus comparing the intensities obtained from amplitude and phase squeezed light, as the polarization of the coherent state component of the driver varies from linear $(A=0$) to circular $(A=1$).~For linear polarization, where the $\perp$-component is in a bright squeezed vacuum (BSV), the squeezing direction has no impact on the HHG spectrum.~In this regime, the process is dominated by the coherent $\parallel$-component, while the BSV contribution barely modifies the HHG spectrum characteristics. In this case, the fluctuations induced by squeezing alter the shot-to-shot ellipticity of the field, lowering the HHG efficiency compared to the linearly polarized case.~Moreover, since BSV light lacks an intrinsic phase reference (unlike DSV states), no distinction arises between amplitude and phase squeezing; all squeezing directions in phase space are effectively equivalent, such that $S_{\phi=0}(\omega) = S_{\phi=\pi}(\omega)$ and $\Delta S_{q,A=0}(\phi) = 0$.

As $A$ increases, however, the coherent displacement along the $\perp$-polarization counterbalances the $\parallel$ contribution, and the HHG process becomes predominantly governed by the squeezing-induced fluctuations.~The outcome then depends strongly on whether amplitude or phase squeezing is applied.~This effect is most visible for harmonics near the cutoff, where $\Delta S_{q,A}(\phi=0)$ increases nearly monotonically. In this region, the effective ellipticity introduced by phase or amplitude squeezing more strongly affects the underlying electron trajectories that give rise to high-harmonic radiation~\cite{rivera-dean_structured_2025}, underscoring the enhanced sensitivity of the cutoff region to non-classical fluctuations.

%% file: Ellipticity.tex
Having characterized the influence of the driving field on the harmonics' intensity, we now analyze the polarization properties of the harmonics themselves, and specifically focus on their ellipticity.~Under entirely classical driving fields, elliptically polarized harmonics can be produced by using moderately elliptically polarized drivers~\cite{antoine_theory_1996}.~However, the rapid decay of the HHG yield with increasing driving-field ellipticity makes the generation of elliptical high harmonics particularly challenging with single-color fields.~Two-color configurations, where each color is circularly polarized with opposite handedness, overcome this limitation by enforcing spin angular momentum conservation from the driver to the harmonics~\cite{fleischer_spin_2014,pisanty_spin_2014}.~Here, in contrast, we find that introducing squeezing in one of the polarization direction enables generation of highly elliptically polarized harmonics even with single-color drivers, providing an alternative route to polarization control of harmonic radiation.

To quantify this, we evaluate the ellipticity of the $q$th harmonic as $\mathsf{E}_q = \lvert\langle \hat{S}_{3,q} \rangle/\langle \hat{S}_{0,q} \rangle\rvert$, where $\hat{S}_{0,q} = \hat{a}^\dagger_{\parallel,q} \hat{a}_{\parallel,q}+ \hat{a}^\dagger_{\perp,q} \hat{a}_{\perp,q}$ and $\hat{S}_{3,q} = i (\hat{a}^\dagger_{\parallel,q} \hat{a}_{\perp,q} - \hat{a}^\dagger_{\perp,q} \hat{a}_{\parallel,q} )$ are the operator forms of the Stokes parameters.~Here, $\langle\hat{S}_{0,q}\rangle$ represents the total intensity of the $q$th harmonic order, while $\langle\hat{S}_{3,q}\rangle$ encodes its polarization handedness.~Thus, $\mathsf{E}_q$ ranges from 0 to 1 as the emission evolves from linear to circular polarization, respectively.

\begin{figure}
    \centering
    \includegraphics[width=1\columnwidth]{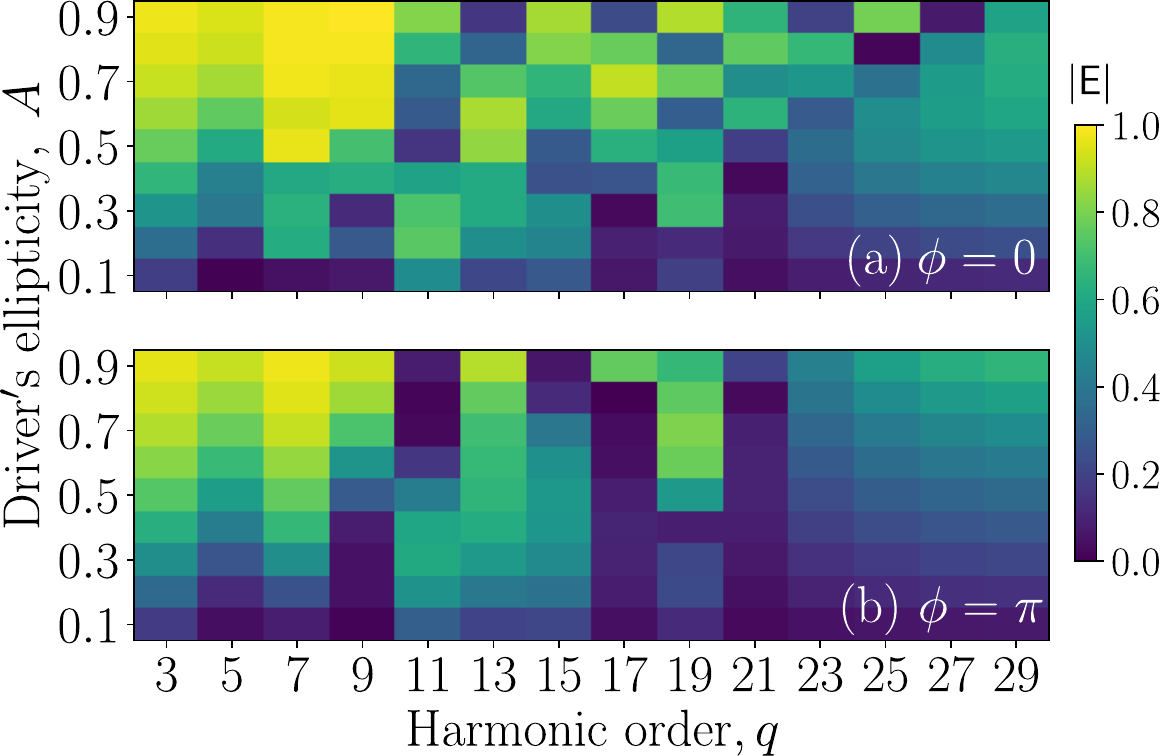}
    \caption{Ellipticity against the harmonic order and the driver's ellipticity in the case of (a) amplitude squeezing and (b) phase squeezing applied to the driving field.~The same conditions as those in Fig.~\ref{Fig:Driver:Charac} are considered here.}
    \label{Fig:HHG:Ellipticity}
\end{figure}

Figure~\ref{Fig:HHG:Ellipticity} displays the ellipticity for different harmonic orders as a function of the driver's ellipticity for both amplitude and phase squeezed driving fields, shown in panels (a) and (b), respectively. Overall, the harmonic ellipticity increases with that of the driving field across the entire spectral range, reaching values as high as $\mathsf{E}_9 = 0.997$ for $A = 0.9$.~However, both squeezing types lead to markedly different behaviors: phase squeezing produces a smoother, more gradual increase of $\mathsf{E}_q$ with $A$, whereas amplitude squeezing yields a more irregular dependence involving a larger number of harmonics, as the cutoff frequency increases with $A$ (see Supplementary Material~\ref{Sec:App:Complementary}). Among both types, amplitude squeezing results in higher average values of $\mathsf{E}_q$.

These features can be understood from how field fluctuations of the driver map onto the mean intensity of the polarization components of the generated harmonics at high values of $A$ (see Supplementary Material~\ref{Sec:App:Complementary}).~For amplitude-squeezed drivers, the mean intensities of both the $\parallel$ and $\perp$ components of the $q$th harmonic are nearly identical, but the stronger phase noise induced by the field fluctuations prevents them from achieving $\mathsf{E}_q = 1$ across all harmonic orders.~In contrast, for phase squeezed drivers the intensity along the squeezed polarization components is overall higher than in the orthogonal one due to the enhanced field fluctuations, which hinders the condition for circular polarization---equal intensity and a $\pi/2$ phase shift between components. These features, however, predominantly affect the higher harmonic orders which are more sensitive to squeezing, while the lower ones ($q\in[3,9]$) behave similarly for both squeezing types.

%% file: PhotonStatistics.tex
When coherent state light sources are used, the approaches in Refs.~\cite{antoine_theory_1996,fleischer_spin_2014,pisanty_spin_2014} lead to the generation of perfectly coherent, elliptically polarized harmonics, provided that the driving field strength is low enough to avoid significant depletion of the atomic system~\cite{stammer_entanglement_2024}.~Thus, under these conditions the generated harmonics exhibit Poissonian photon statistics~\cite{lewenstein_generation_2021,rivera-dean_strong_2022,stammer_quantum_2023}. In contrast, the configurations studied here present a markedly different scenario:~the generation of highly elliptically polarized harmonics requires not only large driving-field ellipticities but also sufficient field fluctuations introduced by squeezing.~In the absence of such squeezing-induced fluctuations, harmonic generation is suppressed~\cite{rivera-dean_structured_2025}, indicating that the emitted harmonics originate from the field fluctuations of the driver.~Consequently, their photon statistics are expected to deviate substantially from those obtained when using coherent state drivers~\cite{lemieux_photon_2024,tzur_measuring_2025,stammer_weak_2025,rivera-dean_attosecond_2025}.

To quantitatively characterize the photon statistics of the harmonics, we use the second-order autocorrelation function, $g^{(2)}(0) = \langle \hat{a}^{\dagger 2} \hat{a}^2 \rangle/\langle \hat{a}^\dagger \hat{a}\rangle ^2$, which in our case can be explicitly written for a given optical mode $(\mu,q)$ as~ (see Supplementary Material~\ref{Sec:SM:TheoryBackground})
\begin{equation}
	g^{(2)}_{\mu,q}(0) 
		 = \frac{\int \dd \Tilde{\varepsilon}^{(\perp)}_{\alpha}
		 	Q_{\perp}(\Tilde{\varepsilon}^{(\perp)}_{\alpha})
	 		|d^{(\mu)}_{\boldsymbol{\varepsilon}_{\alpha}}(\omega)|^4}{\big[\int \dd \Tilde{\varepsilon}^{(\perp)}_{\alpha}
	 		Q_{\perp}(\Tilde{\varepsilon}^{(\perp)}_{\alpha})
	 		|d^{(\mu)}_{\boldsymbol{\varepsilon}_{\alpha}}(\omega)|^2\big]^2}.
\end{equation}
Here, $g_{\mu, q}^{(2)}(0)$ quantifies photon-number correlations within the $(\mu,q)$ optical mode: values greater, equal or lower than unity correspond to super-Poissonian, Poissonian and sub-Poissonian statistics, respectively.~The latter case constitutes a sufficient, though non-necessary, condition for identifying a field as non-classical~\cite{LoudonBookCh5}.

\begin{figure}
	\centering
	\includegraphics[width=1\columnwidth]{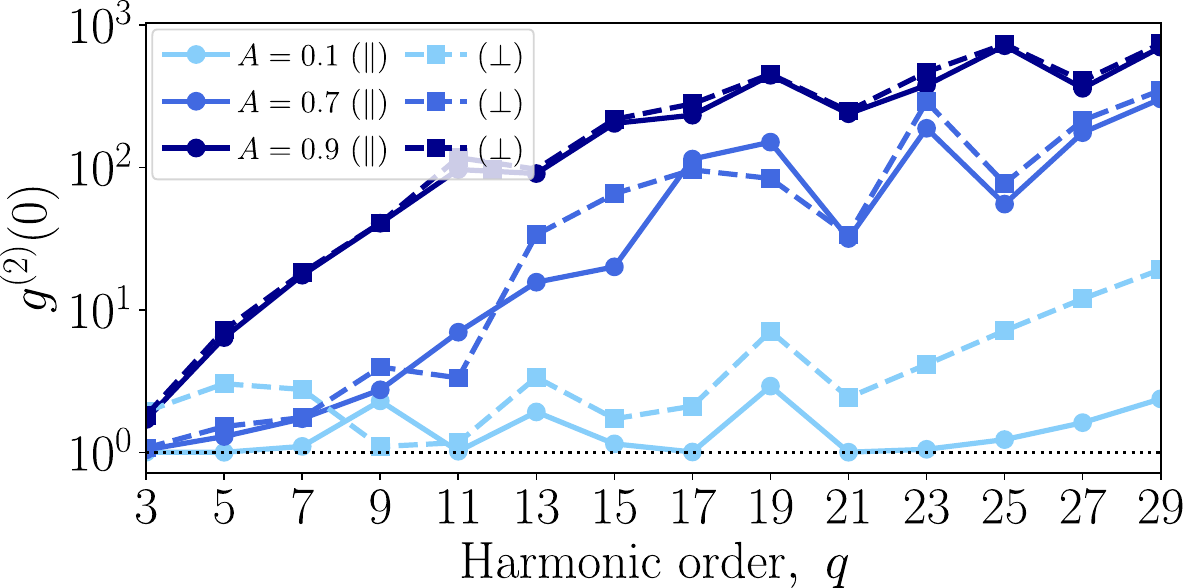}
	\caption{Second-order autocorrelation function $g_{\mu,q}^{(2)}(0)$ as a function of the harmonic order $q$, with the colors indicating the driving field ellipticity. Solid (dashed) curves correspond to the $\mu = \parallel$ ($\mu = \perp$) polarization component, with the different colors denoting.~The dotted thin black line marks the value $g^{(2)}(0)=1$.~The same conditions as those in Fig.~\ref{Fig:Driver:Charac} are used here.}
	\label{Fig:g2:Amp:squ}
\end{figure}

Figure~\ref{Fig:g2:Amp:squ} displays the $g_{\mu,q}^{(2)}(0)$ function for several harmonic orders under amplitude-squeezed driving fields, with similar trends obtained for the phase-squeezed case. In all situations, $g_{\mu,q}^{(2)}(0) > 1$, although its magnitude strongly depends on the initial polarization configuration. For almost linearly polarized drivers ($A = 0.1$), the harmonics exhibit moderate super-Poissonian statistics with $1 < g_{\mu,q}^{(2)} < 10$ for harmonics beyond the cutoff. In contrast, for large driving field ellipticities, $g_{\mu,q}^{(2)}(0)$ grows exponentially, reaching values as high as $10^{3}$ for nearly circular polarization. A similar exponential enhancement also occurs for small ellipticities, but only for harmonics beyond the cutoff region.

This exponential growth of $g_{\mu,q}^{(2)}(0)$ originates from the conditions under which the harmonics are generated. As discussed earlier, in the absence of squeezing features, large ellipticities in the driving field suppress harmonic generation, meaning that the rare, high-intensity events within the driver's probability distribution $\mathcal{Q}(\varepsilon)$ dominate the emission in the presence of squeezing.~Moreover, HHG is a strongly nonlinear process, with the plateau harmonics' intensity scaling as $I_q \propto \abs{d(\omega_q)}^2 \propto I_L^p$ ($p>0$)~\cite{kulander_theory_1990,lhuillier_calculations_1992,weissenbilder_how_2022}. The interplay between this pronounced nonlinearity and the broad field fluctuations characteristic of strongly squeezed light leads to the extremely large values of $g_{\mu,q}^{(2)}(0)$ observed for almost circular polarized drivers---or, equivalently, for linearly polarized BSV fields (see Supplementary Material~\ref{Sec:App:g2:further:toy}). Nonetheless, in regimes where a coherent component can already sustain harmonic generation, for instance for $A = 0.1$ in Fig.~\ref{Fig:g2:Amp:squ}, the introduction of squeezing mainly adds fluctuations around the coherent baseline, similarly to what has been recently observed in bichromatic HHG schemes where one component exhibits squeezing features~\cite{lemieux_photon_2024,tzur_measuring_2025,stammer_weak_2025,rivera-dean_attosecond_2025}. However, even in the absence of such coherent baseline, for extremely large squeezing levels depletion of the driving field---leading to a reduced conversion efficiency~\cite{rivera-dean_propagation_2025}---is expected to prevent the unbounded growth of $g_{\mu,q}^{(2)}(0)$ (see Supplementary Material \ref{Sec:App:g2:further:depletion}).

%% file: Supplementary_Material.tex
\newpage
\onecolumngrid
\begin{center}
    {\large \textbf{\textsc{Supplementary Material}}}
\end{center}

\tableofcontents

\section{Theory background}\label{Sec:SM:TheoryBackground}
In this work, we consider the interaction of a high-intensity elliptically polarized field exhibiting squeezing features with an atomic system initially in its ground state. The initial quantum state of the joint light-matter system can be expressed as
\begin{equation}
	\ket{\Psi(t_0)}
		 = \ket{\text{g}}
			 \otimes \ket{\bar{\alpha}_{\parallel}}
			 	 \otimes \hat{D}(\bar{\alpha}_{\perp})\hat{S}(\xi)\ket{0}
			 	 	\otimes \ket{\{0\}_{q>1}},
\end{equation}
where $\hat{D}(\alpha) = \exp[\alpha\hat{a}^\dagger - \alpha^*\hat{a}]$ is the displacement operator, and $\hat{S}(\xi) = \exp[\xi^*\hat{a}^2 - \xi \hat{a}^{\dagger 2}]$ is the squeezing operator, with $\xi = r e^{-i\phi}$ ($r>0$) denoting the squeezing parameter. To describe elliptically polarized light, we set $\bar{\alpha}_{\perp} = -iA \lvert\bar{\alpha}_{\parallel}\rvert$, with $0\leq A\leq 1$ defines the ellipticity of the driving field: when $A = 0$ the input field is linearly polarized, whilst for $A=1$ it is circularly polarized. All harmonic modes other than the fundamental ($q=1\equiv L$) are initially in the vacuum state. In the density matrix formulation, the initial state can thus be written as
\begin{equation}
	\hat{\rho}(t_0)
		= \dyad{\text{g}} \otimes \hat{\rho}_{\text{driver}}(t_0) \otimes \dyad{\{0\}_{q>1}}.
\end{equation}

For convenience, in this work, we represent the quantum state of the driving field using the generalized positive-$P$ representation~\cite{drummond_generalised_1980}
\begin{equation}
	\hat{\rho}_{\text{driver}}(t_0)
		= \bigotimes_{\mu = (\perp,\parallel)}
			\int \dd^2 \alpha_{\mu}
				\int \dd^2 \beta_{\mu}
							\dfrac{P_{\mu}(\alpha_{\mu},\beta_{\mu}^*)}{\langle \beta_{\mu}^*\vert \alpha_{\mu}\rangle}
								\dyad{\alpha_{\mu}}{\beta_{\mu}^*},
\end{equation}
where $P_\mu(\alpha_\mu, \beta_\mu^*)$ is chosen as a positive-definite function given by~\cite{d_drummond_quantum_2016}
\begin{equation}
	P_{\mu}(\alpha_\mu,\beta_\mu^*)
		= \dfrac{1}{4\pi}
				\exp[-\dfrac{\abs{\alpha_\mu - \beta_\mu^*}^2}{4}]
					Q_\mu\bigg(\dfrac{\alpha_\mu +\beta_\mu^*}{2}\bigg),
\end{equation}
with $Q_\mu(\alpha) = \pi^{-1} \mel{\alpha}{\hat{\rho_{\mu}}}{\alpha}$ denoting the Husimi function of the state $\hat{\rho}_{\mu}$~\cite{SchleichBookCh12}.

Within this framework, the quantum state after strong laser-matter interaction with an atomic system can be expressed as~\cite{gorlach_high-harmonic_2023,rivera-dean_structured_2025}
\begin{equation}\label{Eq:SM:final:state}
	\begin{aligned}
	\hat{\rho}(t)
		&= \int \dd^2 \alpha_{\perp}
				\int \dd^2 \beta_{\perp}
					\int \dd^2 \alpha_{\parallel}
						\int \dd^2 \beta_{\parallel}
							\dfrac{P_{\perp}(\alpha_{\perp},\beta_{\perp}^*)}{\langle \beta_{\perp}^*\vert \alpha_{\perp}\rangle}
								\dfrac{P_{\parallel}(\alpha_{\parallel},\beta_{\parallel}^*)}{\langle \beta_{\parallel}^*\vert \alpha_{\parallel}\rangle}
						\\&\hspace{3cm}\times
						\dyad{\phi_{\boldsymbol{\alpha}}(t)}{\phi_{\boldsymbol{\beta}^*}(t)}
								\big[
									\hat{D}_{\perp}(\alpha_\perp)
									\otimes \hat{D}_{\parallel}(\alpha_{\parallel})
								\big]
								\bigotimes_{q=1}
									\dyad{\chi_{\boldsymbol{\alpha},q}(t)}{\chi_{\boldsymbol{\beta}^*,q}(t)}
								\big[
									\hat{D}^\dagger_{\perp}(\alpha_\perp)
									\otimes \hat{D}^\dagger_{\parallel}(\alpha_{\parallel})
								\big],
	\end{aligned}
\end{equation}
where we denote $\boldsymbol{\alpha} = (\alpha_{\perp},\alpha_{\mu})$, and define $\relaxket{\chi_{\boldsymbol{\alpha},q}(t)} \equiv \relaxket{\chi^{(\perp)}_{\boldsymbol{\alpha},q}(t)}\otimes\relaxket{\chi^{(\parallel)}_{\boldsymbol{\alpha},q}(t)}$, with $\chi_{\alpha_{\perp},q}(t)$ the amplitude of the $q$th harmonic order in polarization component $\mu$~\cite{lewenstein_generation_2021,rivera-dean_strong_2022,stammer_quantum_2023}
\begin{equation}
	\chi^{(\mu)}_{\boldsymbol{\alpha},q}(t)
		\propto
			\boldsymbol{\epsilon}_{\mu}\cdot
			 \int \dd \tau
			 	\mel{\text{g}}{\hat{\boldsymbol{d}}_{\boldsymbol{\alpha}}(\tau)}{\text{g}}
			 		e^{-i\omega_q \tau} , 
\end{equation}
where $\boldsymbol{d}_{\boldsymbol{\alpha}}(t)$ is the time-dependent dipole moment operator under the influence of $\boldsymbol{E}_{\boldsymbol{\alpha}}(t) = \mel{\boldsymbol{\alpha}}{[\hat{\boldsymbol{E}}_{L,\perp}(t) + \hat{\boldsymbol{E}}_{L,\parallel}(t)]}{\boldsymbol{\alpha}}$, where $\hat{\boldsymbol{E}}_{q,\mu}(t) = \boldsymbol{\epsilon}_{\mu}[\hat{a}_{q,\mu} e^{-i\omega_q t} + \text{c.c.}]$ is the electric field operator acting on mode $(q,\mu)$. Finally, in Eq.~\eqref{Eq:SM:final:state}, $\ket{\phi_{\boldsymbol{\alpha}}(t)}$ denotes the electronic state, which satisfies
\begin{equation}
	i\hbar \pdv{\ket{\phi_{\boldsymbol{\alpha}}(t)}}{t}
		= 
			\big[
				\hat{H}_{\text{atom}}
				+ \mathsf{e} \hat{r}\cdot \boldsymbol{E}_{\boldsymbol{\alpha}}(t)
			\big]
			\ket{\phi_{\boldsymbol{\alpha}}(t)}.
\end{equation} 

\subsection{Evaluation of observables on the harmonic modes}
Our discussion mainly develops around two observables acting on the harmonic modes: the harmonic spectrum $S(\omega)$ and the second-order autocorrelation function $g^{(2)}(0)$.~Both quantities can be expressed in terms of expectation values of physical observables $\hat{O}_{q,\mu}$, i.e.,
\begin{equation}
	\langle \hat{O}_{q,\mu}\rangle
		= \tr[\hat{O}_{q,\mu}\hat{\rho}(t)],
\end{equation}
where $\hat{O}_q(t)$ denotes some normally ordered combination of creation and annihilation operators acting on the mode $(q,\mu)$.

By inserting Eq.~\eqref{Eq:SM:final:state} into the expression above, and noting that in the low-depletion HHG regime~\cite{stammer_entanglement_2024} one can approximate $\braket{\boldsymbol{\beta}^*}{\boldsymbol{\alpha}} \approx \braket{\phi_{\boldsymbol{\beta}^*}(t)}{\phi_{\boldsymbol{\alpha}}(t)}\prod_q\braket{\boldsymbol{\chi}_{\boldsymbol{\beta}^*,q}(t)}{\boldsymbol{\chi}_{\boldsymbol{\alpha},q}(t)}$~\cite{tzur_generation_2023}, we arrive at
\begin{equation}\label{Eq:SM:mel:O}
	\langle \hat{O}_{q,\mu} \rangle
		= \int \dd^2 \alpha_{\mu}
				\int \dd^2 \beta_{\mu}
					\int \dd^2 \alpha_{\bar{\mu}}
						\int \dd^2 \beta_{\bar{\mu}}
							\dfrac{P_{\mu}(\alpha_\mu,\beta_\mu^*)}{\langle\chi^{(\mu)}_{\boldsymbol{\beta}^*,q}(t)\vert\chi^{(\mu)}_{\boldsymbol{\alpha},q}(t)\rangle}
							P_{\bar{\mu}}(\alpha_{\bar{\mu}},\beta_{\bar{\mu}}^*)
							\langle\chi^{(\mu)}_{\boldsymbol{\beta}^*,q}(t)\vert\hat{O}_{q,\mu}\vert\chi^{(\mu)}_{\boldsymbol{\alpha},q}(t)\rangle.
\end{equation}

To evaluate these observables, we work in the classical limit, also referred to as the \emph{small single photon amplitude} limit in Ref.~\cite{gorlach_high-harmonic_2023}.~In this regime, coherent state amplitudes are related to the field amplitude via $\varepsilon_{\alpha} = 2 \epsilon \alpha$, with $\epsilon \propto 1/\sqrt{V}$ and $V$ the quantization volume.~Taking the limit $\epsilon \to 0$ ($V\to\infty$), which is justified for free-space propagating fields~\cite{tannoudji_classical_1997}, requires simultaneously sending $\alpha \to \infty$ to ensure a finite field strength $\varepsilon_{\alpha}$.~This second limit is consistent with our problem, since we deal with driving fields of extremely large mean photon numbers.~Under this limit, Eq.~\eqref{Eq:SM:mel:O} becomes
\begin{equation}\label{Eq:SM:obs:cl:lim}
		\langle \hat{O}_{q,\mu} \rangle
	= \int \dd^2 \varepsilon_{\alpha,\mu}
			\int \dd^2 \varepsilon_{\alpha,\bar{\mu}}
				\lim_{\epsilon\to0}
					\Big[
						\dfrac{1}{16\epsilon^4} Q_{\mu}(\varepsilon_{\alpha_\mu})
							Q_{\bar{\mu}}(\varepsilon_{\alpha_{\bar{\mu}}})
					\Big]
					\langle\chi^{(\mu)}_{\boldsymbol{\alpha},q}(t)\vert\hat{O}_{q,\mu}\vert\chi^{(\mu)}_{\boldsymbol{\alpha},q}(t)\rangle,
\end{equation}
where in this case a Dirac delta $\delta(\varepsilon_{\alpha,\mu}-\varepsilon_{\beta,\mu})$ arises naturally from the evaluated limit~\cite{gorlach_high-harmonic_2023,rivera-dean_structured_2025}.

\begin{figure}
    \centering
    \includegraphics[width=1\textwidth]{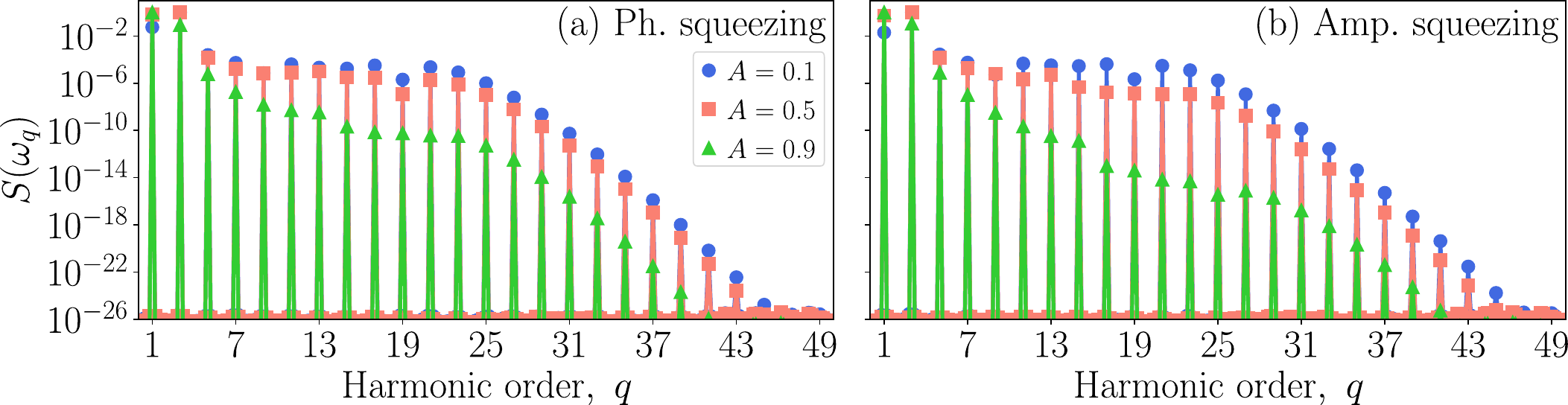}
    \caption{High harmonic generation spectrum for (a) phase and (b) squeezed drivers of different ellipticities $A$.~Calculations have been performed over five optical cycles of a monochromatic field with set $\overline{\varepsilon}_\parallel = 0.053$ a.u., $\omega_L = 0.057$ a.u., $I_{\text{sq},\perp} = 10^{-5}$ a.u., using an ionization potential $I_p = 0.5$ a.u.~(hydrogen).}
    \label{Fig:SM:Spectrum}
\end{figure}

It is important to stress that in the expression the limit does not affect the expectation value directly. In general, however, one should be cautious, since $\langle\chi^{(\mu)}_{\boldsymbol{\alpha},q}(t)\vert\hat{O}_{q,\mu}\vert\chi^{(\mu)}_{\boldsymbol{\alpha},q}(t)\rangle \propto \epsilon^n$, with $n$ an integer that depends on the observable under consideration. Thus, extra care must be taken when evaluating the limit~\cite{stammer_weak_2025}. For the quantities considered here---$S(\omega)$ and $g^{(2)}(0)$---this issue does not arise, as they are defined by 
\begin{equation}
	S(\omega)
		\propto \pdv{}{\omega_q}
			\bigg[
				\sum_{q\neq 1,\mu} 
					\hbar \omega_q
						\langle\hat{a}_{q,\mu}^\dagger \hat{a}_{q,\mu}\rangle
			\bigg],
	\quad
	g^{(2)}(0)
		= \dfrac{\langle \hat{a}^{\dagger2}_{q,\mu}\hat{a}^2_{q,\mu}\rangle}{\langle \hat{a}^{\dagger}_{q,\mu}\hat{a}_{q,\mu}\rangle},
\end{equation}
which by construction the dependence on $\epsilon$ cancels, either due to the continuous frequency in $S(\omega)$ or the structure of $g^{(2)}(0)$ itself. Consequently, the limits $\lim_{\epsilon\to0} [S(\omega)]$ and $\lim_{\epsilon\to0}[g^{(2)}(0)]$ remain well-defined. Figure~\ref{Fig:SM:Spectrum} displays some examples for $S(\omega)$ for (a) phase and (b) amplitude squeezed drivers, with the different colors and marker-styles denoting different driving field ellipticities.

\subsection{Change of coordinates}
The evaluation of the limits presented in Eq.~\eqref{Eq:SM:obs:cl:lim} has been extensively discussed in the literature~\cite{gorlach_high-harmonic_2023,rivera-dean_structured_2025}. However, most studies to date have restricted to the cases of amplitude and phase squeezing, i.e., situations where $\xi \in \mathbbm{R}$. Accounting for the intrinsically complex nature of $\xi$ requires some care in order to properly recover expressions analogous to those reported in the aforementioned works.

On the one hand, for the coherent state component it was found that~\cite{rivera-dean_structured_2025}
\begin{equation}
	\lim_{\epsilon\to 0}
		\bigg[
			\dfrac{1}{4\epsilon^2}Q_{\parallel}(\varepsilon_{\parallel})
		\bigg]
		= \delta(\varepsilon_{\parallel,x} - \bar{\varepsilon}_x)
			\delta(\varepsilon_{\parallel,y} - \bar{\varepsilon}_y),
\end{equation}
where we have decomposed $\varepsilon_{\parallel} = \varepsilon_{\parallel,x} + i \varepsilon_{\parallel,y}$, and introduced $\bar{\varepsilon}_{\parallel} = 2 \epsilon \bar{\alpha}_{\parallel}$.~On the other hand, for a squeezed state this evaluation requires more care. In general, the Husimi function of a squeezed coherent state with coherent state amplitude $\bar{\alpha}$ can be written as
\begin{equation}
	Q(\alpha)
		= \dfrac{1}{\cosh(r)}
				\exp[-\abs{\tilde{\alpha}}^2
						- \dfrac{\tanh(r)}{2}
							\big(
								(\Tilde{\alpha})^2
								+ 	(\Tilde{\alpha}^*)^{2}
							\big)
						],
\end{equation}
where we define $\Tilde{\alpha} = (\alpha - \bar{\alpha})e^{-i\phi/2}$. Expanding the exponent into its real and imaginary components, i.e., $\Tilde{\alpha} = \tilde{\alpha}_x + i \tilde{\alpha}_y$, the Husimi function an be rewritten as
\begin{equation}
	Q(\alpha)
		= \dfrac{1}{\cosh(r)}
			\exp[- \dfrac{\Tilde{\alpha}^2_x}{1+e^{2r}}
				-  \dfrac{\Tilde{\alpha}^2_y}{1+e^{-2r}}],
\end{equation}
with the variables $(\Tilde{\alpha}_x,\tilde{\alpha}_y)$ related to the original integration variables through
\begin{equation}
	\begin{aligned}
	&\Tilde{\alpha}_x
		= (\alpha_x-\bar{\alpha}_x) \cos(\tfrac{\phi}{2})
			- (\alpha_y-\bar{\alpha}_y) \sin(\tfrac{\phi}{2}),
	\quad\quad
	\Tilde{\alpha}_y
		= (\alpha_y-\bar{\alpha}_y)\cos(\tfrac{\phi}{2})
		+(\alpha_x-\bar{\alpha}_x) \sin(\tfrac{\phi}{2}).
	\end{aligned}
\end{equation}

Thus, this change of variables effectively shifts the phase-space coordinate system to the squeezing origin and rotates such that the state is represented as a BSV state squeezed along the optical quadrature defined by $\tilde{\alpha}_y$. In this coordinate system, applying the classical limit becomes straightforward and yields
\begin{equation}\label{Eq:SM:lim:function}
	\lim_{\epsilon\to 0}
		\bigg[
			\dfrac{1}{4\epsilon^2}Q_{\perp}(\varepsilon_{\perp})
		\bigg]
			= \dfrac{1}{\sqrt{8\pi I_{\text{squ}}}}
					\exp[- \dfrac{\Tilde{\varepsilon}_{\perp,x}}{8I_{\text{squ}}}]
						\delta(\varepsilon_{\perp,x} - \bar{\varepsilon}_x),		
\end{equation}
with the subsequent analysis proceeding analogously Refs.~\cite{gorlach_high-harmonic_2023,rivera-dean_structured_2025} once the change of variables has been introduced.

\subsection{Numerical analysis}
In our numerical analysis, we set $\bar{\varepsilon}_{\parallel,x} = \bar{\varepsilon}_{\perp,y} = 0.053$ a.u.~and $\bar{\varepsilon}_{\parallel,y} = \bar{\varepsilon}_{\perp,x} = 0.0$ a.u., with $I_{\text{squ}} = 10^{-5}$ a.u. for the squeezing intensity.~Using these parameters, the spectral amplitudes $\boldsymbol{\chi}_{\boldsymbol{\alpha},q}(t)$ were numerically computed within the strong-field approximation (SFA)~\cite{lewenstein_theory_1994} using the \texttt{RB-SFA} Mathematica package~\cite{RBSFA}. Specifically, 140 points along the Husimi distribution were sampled for each value of $A$ and $\theta$ considered in the main text, and subsequently used for the analysis of both $S(\omega)$ and $g^{(2)}(0)$. The calculations were performed with a monochromatic driving field extended over five optical cycles. As the atomic system, we considered hydrogen ($I_p = 0.5$ a.u.) with $\omega = 0.057$ a.u.~for the driving field frequency.

\section{Complementary plots}\label{Sec:App:Complementary}
\begin{figure}[h!]
    \centering
    \includegraphics[width=1\columnwidth]{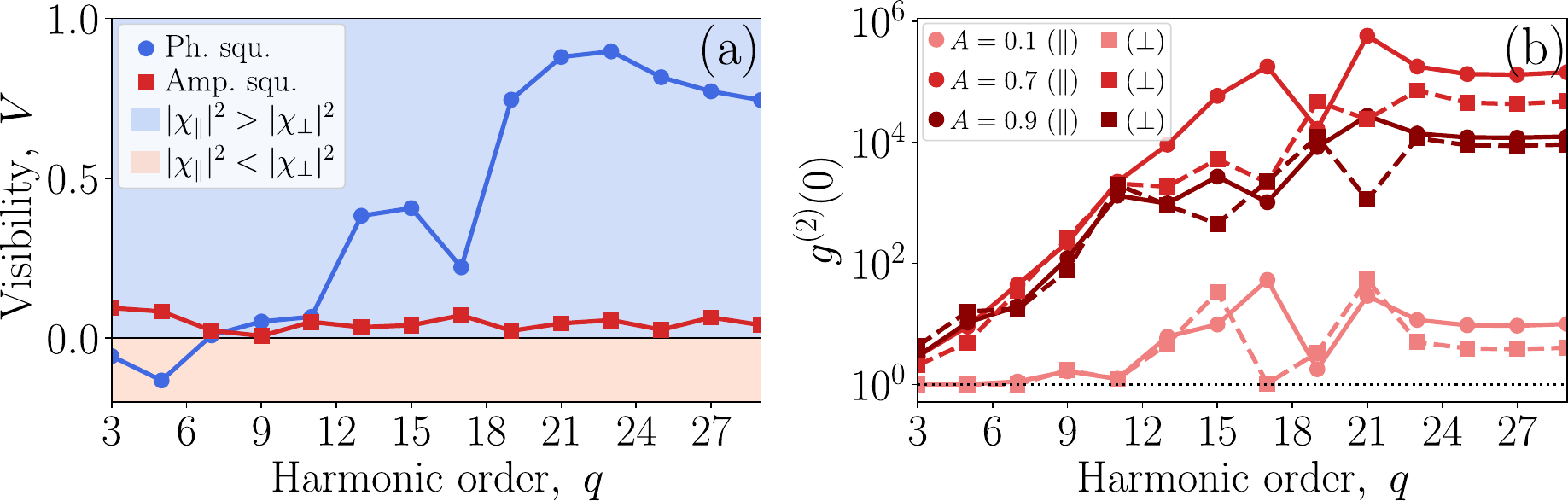}
    \caption{(a) Visibility for different driving field conditions and (b) the 2nd order autocorrelation function $g^{(2)}(0)$ against the harmonic order $q$ for different ellipticity values $A$ of a phase-squeezed driver.}
    \label{Fig:SM:Res:Ph:Squ}
\end{figure}

In this section, we present additional figures that complement those shown in the main text. Figure~\ref{Fig:SM:Res:Ph:Squ}~(a) displays the visibility, defined as
\begin{equation}
    V = \dfrac{\langle \hat{a}^\dagger_{q,\parallel}\hat{a}_{q,\parallel}\rangle - \langle \hat{a}^\dagger_{q,\perp}\hat{a}_{q,\perp}\rangle}{\langle \hat{a}^\dagger_{q,\parallel}\hat{a}_{q,\parallel}\rangle + \langle \hat{a}^\dagger_{q,\perp}\hat{a}_{q,\perp}\rangle},
\end{equation}
for a driver with $A = 0.9$. The purpose of this quantity is to characterize how the intensity is distributed, on average, between the polarization components of the generated harmonics. This allows us to gain insight into the mechanisms leading to $\mathsf{E}_q < 1$ for different types of drivers (phase- or amplitude-squeezed), given that circularly polarized light satisfies $V = 0$ with a phase difference of $\pi/2$ between components.

For the phase-squeezed case (blue curve with circular markers), we find $V> 0$ for harmonic orders within the plateau region, indicating that the main factor preventing perfect circular polarization is how the field fluctuations present in the driver translate into higher intensities along the $\parallel$-component. In contrast, for amplitude squeezing (red curve with square markers), we find $V \approx 0$, suggesting that the deviation from $\mathsf{E}_q = 1$ arises instead from phase fluctuations that induce relative phase shifts between polarization components different from $\pi/2$. In both cases, however, the lower harmonic orders exhibit $V=0$, with phase difference close to $\pi/2$, as these result in high ellipticities in the emitted radiation.

Figure~\ref{Fig:SM:Res:Ph:Squ}~(b) shows the $g^{(2)}(0)$ function for phase-squeezed drivers. Qualitatively, the results are similar to those obtained for amplitude squeezing: $g^{(2)}$ increases with the harmonic order for large $A$, and decreases as $A$ is reduced. Quantitatively, however, the magnitudes differ significantly---phase squeezing yields $g^{(2)}(0)$ values up to three orders of magnitude higher than those observed for amplitude squeezing. This arises from the fact that phase squeezing induces larger fluctuations in the field strength. As discussed below, the combination of this enhancement with the strong nonlinearity of the HHG process accounts for the extremely large $g^{(2)}(0)$ values observed.

\section{Analysis of the second-order autocorrelation function}\label{Sec:App:g2:further}
In the main text, we found that the values of $g^{(2)}(0)$ function exceeded by many orders of magnitude those typically observed for standard super-Poissonian sources, which are usually around $g^{(2)}(0) \approx 2$. The main objective of this section is to justify these unusually large values, which arise from a combination of two factors:
\begin{itemize}
\item the unique scaling of the HHG yield with the intensity of the driving field,
\item the characteristics of the field fluctuations associated with squeezed states at intensities high enough to drive HHG in gaseous systems.
\end{itemize} 

For simplicity in the calculations presented here we restrict our calculations to linearly polarized driving fields and various types of sources, ranging from bright squeezed vacuum states (BSV) to general DSV states with non-vanishing field amplitude. This situation is analogous to moving from linear to circular polarization, since in the regime of vanishing squeezing, the limiting cases of either circularly polarized or BSV sources do not generate HHG radiation.

\subsection{Toy model}\label{Sec:App:g2:further:toy}
Under strong-squeezing regimes, the properties of the harmonic radiation were found to be compatible with those of the mixed state~\cite{gorlach_high-harmonic_2023,tzur_generation_2023,rivera-dean_structured_2025,stammer_weak_2025}
\begin{equation}
	\hat{\rho}_q
	= \int \dd \varepsilon_{\alpha} 
	\mathcal{Q}(\varepsilon_{\alpha},\bar{\varepsilon}_{\alpha})
	\dyad{\bar{\chi}_q(\varepsilon_\alpha)},
\end{equation}
where $\mathcal{Q}(\varepsilon_{\alpha},\bar{\varepsilon}_{\alpha})$ corresponds to the limiting appearing in Eq.~\eqref{Eq:SM:lim:function}. It is important to note that, in writing this expression, additional considerations such as the thermodynamic limit~\cite{stammer_weak_2025} must be taken into account. Specifically, we assume that as as $V\to\infty$, then the number of atoms in the HHG interaction region $N_{\text{at}}\to\infty$, such that $\varrho = N_{\text{at}}/\sqrt{V}$ remains constant. Accordingly, we define $\bar{\chi}_q(\varepsilon_\alpha) \equiv \varrho \int \dd \tau \langle \text{g}\vert \hat{d}_{\varepsilon_\alpha}(\tau)\vert \text{g}\rangle e^{-i\omega_q t}$.

To gain insights into the properties of this state, we introduce a simple toy model where $\bar{\chi}_q(\varepsilon_\alpha) = \varepsilon_{\alpha}^p$. This choice is motivated by the fact that the HHG intensity, particularly in the plateau region, typically scales as a power $p>0$ of the driving field~\cite{kulander_theory_1990,lhuillier_calculations_1992,weissenbilder_how_2022}. Within this model, the corresponding state can be written as
\begin{equation}\label{Eq:SM:toy:model}
	\hat{\rho}_q
	= \dfrac{1}{\sqrt{2\pi\sigma}}
	\int \dd\varepsilon_\alpha
	\exp[-\dfrac{(\varepsilon_\alpha - \bar{\varepsilon})^2}{2\sigma}]
	\dyad{\varepsilon_{\alpha}^p}.
\end{equation}
In this framework setting $\bar{\varepsilon} = 0$ reproduces the analogue of a BSV-driven process, while $\bar{\varepsilon}\neq 0$ corresponds to a general DSV state.~Without loss of generality, we take the integration contour along the real axis, so that the $g^{(2)}(0)$ function becomes
\begin{equation}\label{Eq:SM:g2}
	g^{(2)}(0)
	= \sqrt{2\pi\sigma}\dfrac{\int \dd \varepsilon_\alpha \exp[-(\varepsilon_\alpha - \bar{\varepsilon})^2/2\sigma] \varepsilon_{\alpha}^{4p}}{\big[\int \dd \varepsilon_\alpha \exp[-(\varepsilon_\alpha - \bar{\varepsilon})^2/2\sigma] \varepsilon_{\alpha}^{2p}\big]^2}.
\end{equation}
From the Cauchy-Schwarz inequality, it immediately follows that $g^{(2)}\geq 1$. More explicitly, the denominator can be lower bounded as
\begin{equation}
	\langle \hat{a}^\dagger_q\hat{a}_q\rangle^2
	=
	\bigg[
	\int \dd \varepsilon_{\alpha}
	\mathcal{Q}(\varepsilon_{\alpha})
	\alpha^{2p}
	\bigg]^2
	\leq
	\bigg[
	\int \dd \varepsilon_{\alpha}
	\mathcal{Q}(\varepsilon_{\alpha})
	\alpha^{4p}
	\bigg]
	\bigg[
	\int \dd \varepsilon_{\alpha}
	\mathcal{Q}(\varepsilon_{\alpha})
	\bigg]
	= \int \dd \varepsilon_{\alpha}
	\mathcal{Q}(\varepsilon_{\alpha})
	\alpha^{4p}
	= \langle \hat{a}^{\dagger2}_q\hat{a}^2_q\rangle,
\end{equation}
where we have used the fact that $\mathcal{Q}(\varepsilon_{\alpha})$ is normalized to unity. 

\begin{figure}
	\centering
	\includegraphics[width=0.8\textwidth]{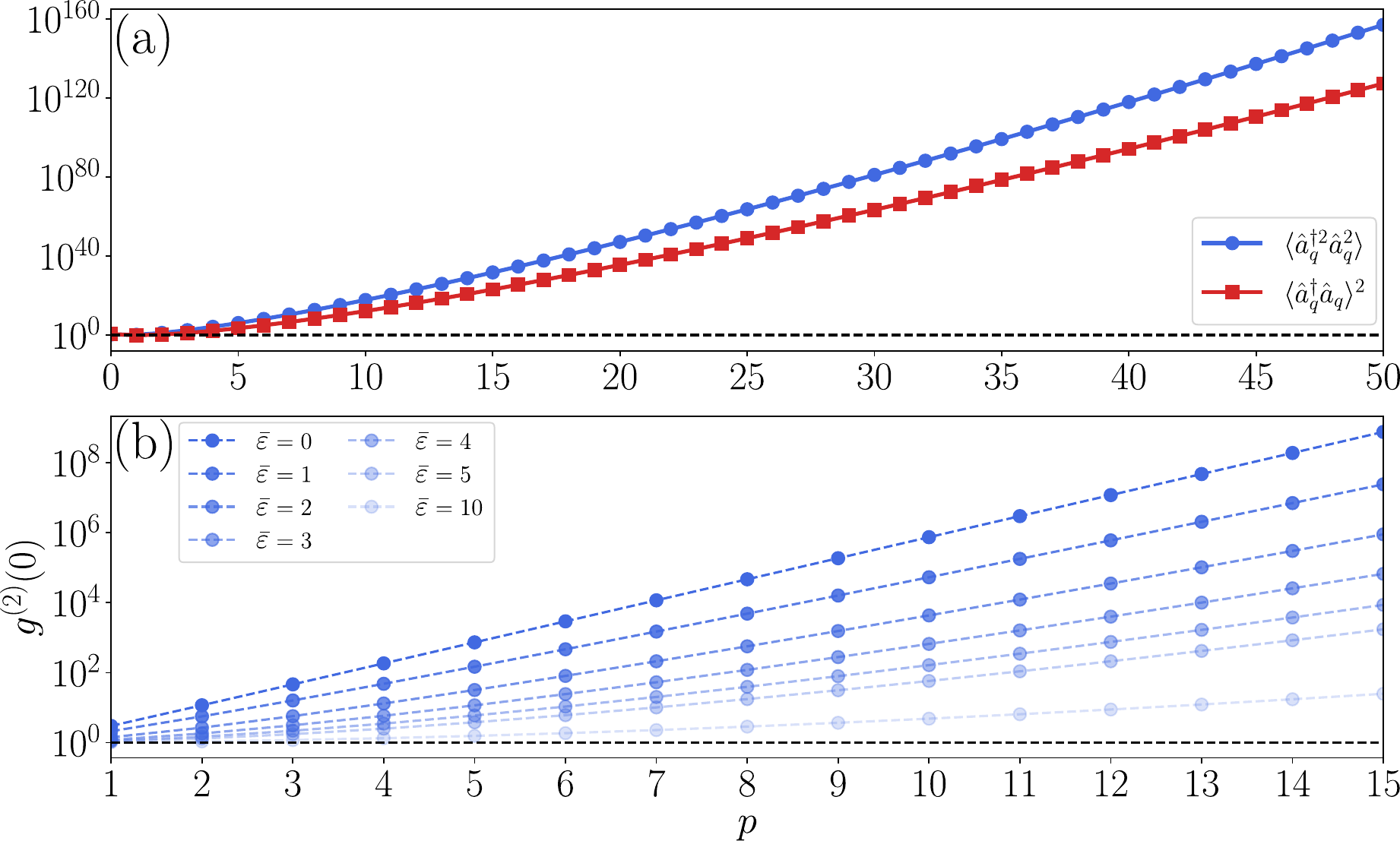}
	\caption{(a) Behavior of $\langle \hat{a}_q^{\dagger 2}\hat{a}_q\rangle$ (blue curve with circular markers) and $\langle \hat{a}_q^{\dagger 2}\hat{a}_q\rangle$ (red curve with circular markers) for the toy-model in Eq.~\eqref{Eq:SM:toy:model} when setting $\bar{\varepsilon} = 0$. (b) Results for various values of $\bar{\varepsilon}$. Here, we have set $\sigma = 1$ for simplicity.}
	\label{Fig:g2:toy:model}
\end{figure}

However, from a more general perspective, explicitly evaluating the integrals in Eq.~\eqref{Eq:SM:g2} for the special case $\varepsilon_{\alpha} = 0$ yields
\begin{equation}
	g^{(2)}_{\text{BSV}}(0)
	= \sigma \sqrt{\pi} \dfrac{\Gamma(\tfrac12 + 2p)}{\Gamma(\tfrac12 + p)^2},
\end{equation}
from which we find that $\lim_{p\to\infty} [g^{(2)}_{\text{BSV}}(0)] = \infty$.~This result is further illustrated in Fig.~\ref{Fig:g2:toy:model}~(a),which shows the behavior of $\langle \hat{a}^{\dagger2}_q\hat{a}^2_q\rangle$ (blue curve) and $\langle \hat{a}^\dagger_q\hat{a}_q\rangle^2$ (red curve) as a function of $p$.~The growing separation between the two curves with increasing $p$ highlights the divergence of $g^{(2)}(0)$.

The important message to extract from this analysis is that the harmonic conversion efficiency counterbalances the rapidly decreasing probability of finding large field amplitudes.~As a consequence, strong intensity correlations emerge in the harmonic emission, since the rare outliers of the distribution become the dominant contributors. In the extreme limit of infinitely large conversion efficiency, the width of the field distribution no longer plays a role in determining the photon statistics of the state, provided it remains greater than zero. In realistic scenarios, however, once the field strength exceeds a critical value, depletion effects start to play a role, potentially causing $g^{(2)}(0)$ to saturate to a constant value.

The more general scenario with $\bar{\varepsilon} \neq 0$ is shown in Fig.~\ref{Fig:SM:g2:depletion}~(b). We observe that increasing $\alpha_0$ reduces the steepness of the $g^{(2)}(0)$ dependence on $p$ (here denoted as $n$). In contrast to the BSV-analogue case, investigating the limit $\sigma \to 0$ yields a finite harmonic intensity, $\abs{\varepsilon}^{2p} \neq 0$.~Thus, the dominant contribution to the HHG spectrum arises from the coherent $\alpha_0$ field. In this limit, $g^{(2)}(0) \to 1$, in stark contrast to the $\varepsilon_0=0$ case, where $g^{(2)}(0)$ becomes ill-defined as $\sigma \to 0$.

\subsection{Introducing depletion of the ground state}\label{Sec:App:g2:further:depletion}
In practice, however, the HHG conversion efficiency does not scale uniformly with the driving field intensity. For coherent states of sufficiently high intensity, ground state depletion becomes a dominant effect, drastically reducing the HHG efficiency. Once depletion is taken into account, the toy model introduced earlier ceases to be valid: above a certain value of $\varepsilon_{\alpha}$, the yield no longer grows exponentially but instead drops sharply.

To investigate how depletion affects the $g^{(2)}$ function, we explicitly include it in the evaluation of the harmonic spectrum.~Our approach follows that of Ref.~\cite{rivera-dean_propagation_2025} which, within the strong-field approximation~\cite{lewenstein_theory_1994}, evaluates the time-dependent dipole moment used for the calculation of $\chi_{q}(\varepsilon_{\alpha}(t)$ as
\begin{equation}\label{Eq:HHG:dipole}
	d(t)
	= \bra{\psi(t)}\hat{d}\ket{\psi(t)}
		\approx 
			\int \dd v \int_{t_0}^t \dd t_1
				e^{-\frac12\int^{t}_{-\infty} \dd \tau \Gamma_{\text{ADK}}(\tau)}
				e^{-iS(p,t,t_1)}
				d\big(p + A(t)\big)
				E(t_1)
				d\big(p + A(t_1)\big)
				e^{-\frac12\int^{t_1}_{-\infty} \dd \tau \Gamma_{\text{ADK}}(\tau)},
\end{equation}
with $S(p,t,t_1)$ is the semiclassical action, $E(t)$ the electric field, and $d(v)$ denotes the transition matrix element between the ground state and a continuum state $\ket{v}$.~The ionization rate $\Gamma_{\text{ADK}}(t)$ is evaluated using the Ammosov-Delone-Krainov (ADK) model~\cite{Ammosov_JETP-ADK_1986}. Numerically, the dipole is computed using the \texttt{RB-SFA} Mathematica package~\cite{RBSFA}, chosen for its efficiency in generating HHG spectra within the SFA. Minor modifications where implemented to incorporate depletion, following the procedure in Ref.~\cite{rivera-dean_propagation_2025}.

\begin{figure}[h!]
	\centering
	\includegraphics[width=1\textwidth]{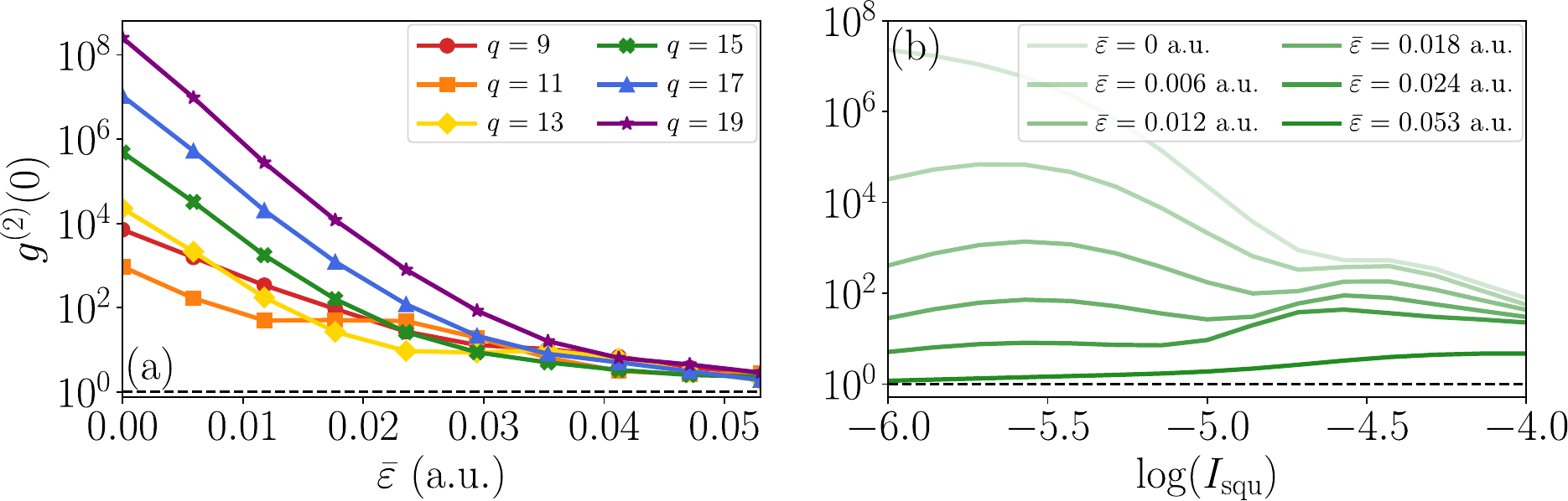}
	\caption{Second-order autocorrelation function as a function of (a) $\bar{\varepsilon}$ for different harmonic orders with fixed $I_{\text{squ}} = 10^{-5}$ a.u.; (b) as a function of $I_{\text{squ}}$ for various values of $\bar{\varepsilon}$ at the 13th harmonic order.~Unlike in the main text, and for simplicity in the treatment of depletion effects, the driving field here is taken with a $\sin^2$ envelope of 13 fs duration and $\omega_L = 0.057$ a.u. for the central frequency.}
	\label{Fig:SM:g2:depletion}
\end{figure}

Figure~\ref{Fig:SM:g2:depletion} shows the values of $g^{(2)}$ when depletion effects are included under various conditions. In panel (a), the squeezing intensity is fixed to $I_{\text{squ}} = 10^{-5}$ a.u., as in the main text, while $\bar{\varepsilon}$ is varied.Although different harmonics exhibit quantitatively distinct values, they all follow the same overall trend as in Fig.~\ref{Fig:g2:Amp:squ}.~More specifically, when $\bar{\varepsilon} = 0.053$ (corresponding to a linearly polarized field with non-vanishing mean amplitude), we find $g^{(2)} \in [1,10]$. As $\bar{\varepsilon} \to 0$—which, in Fig.~\ref{Fig:g2:Amp:squ}, corresponds to moving from linear to circular polarization—$g^{(2)}$ increases exponentially. Furthermore, similar to the case of elliptical polarization, we find that $g^{(2)}$ increases for higher harmonic orders.

Panel (b) instead considers a fixed harmonic order, namely the 13th harmonic (similar behavior is observed for other orders), and shows $g^{(2)}(0)$ as a function of squeezing intensity for several values of $\bar{\varepsilon}$.~As before, the largest values occur when $\bar{\varepsilon} = 0$. In this case, as $I_{\text{squ}}$ increases, depletion effects become more pronounced since the field fluctuations also grow larger.~We observe that beyond approximately $I_{\text{squ}} \approx 5 \times 10^{-5}$ a.u., the value of $g^{(2)}(0)$ saturates and becomes roughly comparable across all $\bar{\varepsilon}$.

From this analysis we conclude that:
\begin{itemize}
	\item the strong nonlinearity of HHG combined with the broad field fluctuations of high-intensity squeezed light are the main factors leading to the large values of $g^{(2)}(0)$;
	\item the effect is most pronounced for BSV or circularly polarized driving fields, since in the absence of field fluctuations HHG would vanish and fluctuations alone sustain the emission;
	\item depletion acts as a stabilizing mechanism, suppressing HHG conversion efficiency once the field amplitude exceeds a critical threshold $\bar{\varepsilon}_{\text{crit}}$, thereby preventing unbounded growth of $g^{(2)}(0)$.
\end{itemize}